\documentclass[a4paper]{article}
    \usepackage[T1]{fontenc}
    \usepackage{graphicx}
    \usepackage{epsfig}
    \usepackage{amsmath}
    \usepackage{amsfonts}
\begin{document}
\fontsize{12}{12} \selectfont

\title{Cosmological scalar fields that mimic the $\Lambda$CDM
cosmological model \footnote{VIII International Conference
"Relativistic Astrophysics, Gravitation and Cosmology" (May 21-23,
2008, Kyiv, Ukraine) Abstracts, p.19}}

\author{V.I. Zhdanov, G. Ivashchenko}
\date{\begin{small}Astronomical Observatory, National Taras
Shevchenko University, Kyiv, Ukraine \\ zhdanov@observ.univ.kiev.ua, g.ivashchenko@gmail.com
\end{small}}
\maketitle

\begin{abstract}
We look for cosmologies with a scalar field (dark energy without
cosmological constant), which mimic the standard $\Lambda$CDM
cosmological model yielding exactly the same large-scale geometry
described by the evolution of the Hubble parameter (i.e.
photometric distance and angular diameter distance as functions of
$z$). Asymptotic behavior of the field solutions is studied in the
case of spatially flat Universe with pressureless matter and
separable scalar field Lagrangians (power-law kinetic term +
power-law potential). Exact analytic solutions are found in some
special cases. A number of models have the field solutions with
infinite behavior in the past or even singular behavior at finite
redshifts. We point out that introduction of the cosmological
scalar field involves some degeneracy leading to lower precision
in determination of $\Omega _m$. To remove this degeneracy
additional information is needed besides the data on large-scale
geometry.
\end{abstract}

\section{Introduction}

\indent\indent Standard $\Lambda$CDM cosmological model explains
the wealth of experimental data on CMB anisotropy and Ia type
supernovae \cite{Apunevych,WMAP,2004}, though some facts require
modification of the dark matter equation of state
\cite{Flores,Ratra} without any revision of the cosmological
constant. On the other hand, theoretical and experimental
developments necessitate  modification of the Standard model
leading to ideas of inflation, cosmological fields, branes in
extra dimensions etc (see, e.g., \cite{Linde,Ratra,Sahni,Shtanov}.
Introduction of scalar fields as a key element of the dark energy
seems to be one of the most simple and natural ways to launch
primordial inflation and to explain the cosmological coincidences
\cite{Linde,SahniStarobinsky}. As soon as the scalar field was
introduced in cosmology, the question about reconstruction of the
field Lagrangian from observational data arose
\cite{AlamSahni,Starobinsky} and a number of interesting examples
have been considered which establish some correspondence between
different models that may be used to explain observational data on
equal terms (see
\cite{AlamSahni,Hoyle,SahniStarobinsky,Saini,Starobinsky}) and
references therein). The solution of this problem appears to be
unstable: small errors in the experimental data lead to
considerable changes in the potential. Introduction of
non-canonical kinetic term like k-essence models
\cite{Bertacca,Chimento,Li,Quersellini,Scherrer} obviously creates
additional ambiguities in reconstruction of the scalar field
Lagrangian.

Introduction  of the additional field may lead to a revision of
some cosmological parameters, even if we retain the large-scale
geometry of the FRW Universe described by the redshift dependence
of the Hubble parameter $H(z)$. Present constraints on
$\Omega_{\Lambda}$,$\Omega_{m}$ within $\sim 2-3{\%}$ accuracy
\cite{Apunevych,WMAP,2004} rely upon measurements of $H(z)$ within
the framework of the $\Lambda$CDM cosmological model, which use
the data on Ia type supernovae magnitude-redshift dependence and
WMAP data that restrict the position of the first peak in the CMB
anisotropy power spectrum. The redshift-space distortions used to
constrain the cosmological parameters from LSS surveys (see, e.g.,
\cite{Alcock,Hoyle}) also have geometrical origin. When additional
degrees of freedom due to the scalar field are introduced, the
same $H(z)$ dependence as in case of the $\Lambda$CDM cosmological
model might be preserved. At the same time this leads to a
reduction of $\Omega_{m}$ if some independent information on this
parameter be used. Of course, there are the other ways to study
the dark matter content (galactic rotation curves, virial mass
estimates, gravitational lensing), however they determine
$\Omega_{m}$ with much lower accuracy. If the scalar field is
introduced leaving somehow the large-scale geometry unchanged, we
cannot separate its contribution into the cosmological density
from the other forms of matter with the same precision as in the
$\Lambda$CDM model.

We keep in mind  this problem in the present paper dealing with
reconstruction of the scalar field Lagrangians that yield exactly
the Hubble diagram of the standard $\Lambda$CDM model. The outline
of the paper is as follows. In Section 2 we write down the basic
relations for spatially flat cosmology with the scalar field and
pressureless matter. In Section 3 we study the problem with
separable scalar field Lagrangians $L=F(S)-V(\varphi)$,
$S=g^{\alpha\beta}\partial_{\alpha}\varphi\partial_{\beta}\varphi/2$
that mimic the Hubble diagram (namely $H(z)$ dependence) of the
$\Lambda$CDM model under different suppositions: with the
canonical kinetic term $F\equiv S$ (Section 3.1); with more
general power-law kinetic term (Section 3.2) or with the linear
and power-law potential $V(\varphi)$ (Section 3.3). Section 4
summarizes the results.

\section {Basic equations}

\indent\indent Here we  present the basic relations of the scalar
field cosmology. Details may be found in
\cite{Bertacca,Li,Sahni,SahniStarobinsky}. We confine ourselves to
the case of the critical cosmological density, which means that
the space-time metric is spatially flat \cite{Alcock}
\begin{equation}\label{eq1}
ds^{2}=dt^{2}-a^{2}(t)[d\chi^{2}+\chi^{2}(d\theta^{2}+\sin^{2}(\theta)d\varphi^{2})],
\end{equation}
($c=1)$, the scale factor $a$ can be related to the redshift $z$:
\begin{equation}\label{eq2}
1+z=a(t_{0})/a(t).
\end{equation}
The observable photometric distance is
\begin{equation*}
 D_{ph}(z)=(1+z)\int\limits_{0}^{z}{[H(\varsigma)]^{-1}}d\varsigma,
\end{equation*}
where $H(z)=\dot{a}(t)/a(t)$ is the  Hubble parameter at the
cosmological epoch $t$. Therefore one may consider $H(z)$ as an
experimentally measurable function yielding $a(t)$ up to an
unessential constant factor.

We consider a cosmological model, where  the main contribution to
the cosmological density is due to the scalar field plus the
matter with zero pressure $p=0$. The mass density $\rho_{m}$ of
the pressureless matter varies as a function of the redshift as
follows:
\begin{equation}\label{eq3}
\rho_{m}=(1+z)^{3}\Omega_{m}\rho_{cr},
\end{equation}
where $\rho_{cr}=3H_{0}^{2}/(8\pi G)$, $\Omega_{m}=\rho_{m}(t_{0})/\rho_{cr}$, $H_{0}=H(0)$ being the modern value of
the Hubble parameter and $t_{0}$ is the modern epoch. The scalar field Lagrangian
\begin{equation}\label{eq4}
L=L(S,\varphi),\quad S=\frac{1}{2}g^{\alpha\beta}\partial_{\alpha}\varphi\partial_{\beta}\varphi
\end{equation}
leads to the Friedmann cosmological  equations, which in case of
spatially flat Universe are
\begin{equation}\label{eq5}
\frac{\ddot{a}}{a}=-\frac{4\pi G}{3}\left[{\rho_{m}+\dot{\varphi}^{2}\frac{\partial L}{\partial S}+2L}\right],
\end{equation}
\begin{equation}\label{eq6}
\left({\frac{\dot{a}}{a}}\right)^{2}=\frac{8\pi G}{3}\left[{\rho_{m}+\dot{\varphi}^{2}\frac{\partial L}{\partial S}-L)}\right],
\end{equation}
\begin{equation*}
S\equiv\dot{\varphi}^{2}/2.
\end{equation*}

Taking (\ref{eq3}) into account we  have $dz/dt=-(1 + z)H(z)$.
This enables us to rewrite (\ref{eq5}),(\ref{eq6}) in terms of
observable quantities $z,H$:
\begin{equation}\label{eq7}
S\frac{\partial L}{\partial S}=\frac{(1+z)}{16\pi G}\frac{dH^{2}}{dz}-\frac{\Omega_{m}}{2}(1+z)^{3}\rho_{cr},
\end{equation}
\begin{equation}\label{eq8}
L(S,\varphi)=\frac{(1+z)^{4}}{8\pi G}\frac{d}{dz}\left({\frac{H^{2}}{(1+z)^{3}}}\right),
\end{equation}
\begin{equation}\label{eq9}
S=\frac{(1+z)^{2}}{2}H^{2}\left({\frac{d\varphi}{dz}}\right)^{2}.
\end{equation}

\section {Scalar field model yielding  Hubble diagram of $\Lambda$CDM}

\indent\indent At present the $\Lambda$CDM cosmological model is in a good agreement with all the observational data having relevance to the space-time geometry. In case of spatially flat cosmological model with pressureless matter (without the scalar field) the cosmological equations yield
\begin{equation}\label{eq10}
H^{2}(z)=H_{0}^{2}h^{2}(z),\quad h(z)\equiv[\Omega_{m}^{0}(1+z)^{3}+1-\Omega_{m}^{0}]^{1/2},
\end{equation}
where the values $H_{0}\approx72km\cdot s^{-1}Mpc^{-1}$, $\Omega_{m}^{0}\approx0.3$ are obtained by fitting the data with the standard $\Lambda$CDM model \cite{WMAP,Apunevych,2004}. Note that in presence of the scalar field the real content $\Omega_{m}$ of the pressureless matter in the cosmological density will be less than $\Omega_{m}^{0}$.

Our first step will be to investigate Eqs.(\ref{eq7})-(\ref{eq9}) in case of the dependence (\ref{eq10}) that will be regarded as the ``observational'' one. We shall present the examples of Lagrangian that lead to the same dependence (\ref{eq10}).

The equations (\ref{eq7})-(\ref{eq9}) may be viewed as observational restrictions on the function $L(S,\varphi)$.
Obviously, they do not fix this function in a unique way. We shall consider less general Lagrangian
\begin{equation}\label{eq11}
L=F(S)-V(\varphi),
\end{equation}
with subsequent specific choice either of the kinetic term $F$ or the potential $V$. The restrictions on these functions on account of (\ref{eq7}),(\ref{eq8}) take on the form
\begin{equation}\label{eq12}
S\frac{dF}{dS}=\frac{\rho_{cr}}{2}(1+z)^{3}(\Omega_{m}^{0}-\Omega_{m}),
\end{equation}
\begin{equation}\label{eq13}
V(\varphi)=F(S)+\rho_{cr}(1-\Omega_{m}^{0}).
\end{equation}

We see from (\ref{eq12}) that for an increasing $F(S)$ one must require that $\Omega_{m}<\Omega_{m}^{0}$: the scalar field ``eats'' part of the cosmological density. For $dF/dS<0$ we have $\Omega_{m}>\Omega_{m}^{0}$. Further we put for definiteness $d\varphi/dz>0$.

\subsection{Quintessence: canonical kinetic term}

\indent\indent For certain known  kinetic term $F(S)$ the field
$\varphi (z)$ may be obtained from (\ref{eq12}); then $V(\varphi
)$ is determined parametrically. Now we proceed to concrete
examples.

In case of the standard kinetic term $F(S) \equiv S$ Eqs.(\ref{eq12}),(\ref{eq13}) are
easily solved to yield
\begin{equation}\label{eq14}
\varphi(z)=\left[{\frac{3(\Omega_{m}^{0}-\Omega_{m})}{8\pi G}}\right]^{1/2}\int\limits_{0}^{z}{d\varsigma}\left[{\frac{(1+\varsigma)}{\Omega_{m}^{0}(1+\varsigma)^{3}+1-\Omega_{m}^{0}}}\right]^{1/2}+\varphi(0),
\end{equation}
\begin{equation}\label{eq15}
V(\varphi )=\frac{\rho_{cr}}{2}\left[{(1+z)^{3}(\Omega_{m}^{0}-\Omega_{m})+2(1-\Omega_{m}^{0})}\right]
\end{equation}
Note that at present epoch $dV/d\varphi\ne0$ ($z = 0$).

For $z\gg1$ we have an exponential growth of $V(\varphi)$
\begin{equation}\label{eq16}
V(\varphi)\approx\frac{\rho_{cr}}{2}(\Omega_{m}^{0}-\Omega_{m})\exp\left\{{\left[{\frac{24G\Omega_{m}^{0}}{\Omega_{m}^{0}-\Omega_{m}}}\right]^{1/2}\varphi}\right\},
\end{equation}
but the field grows only logarithmically as  a function of the
redshift. In case of small deviation of $\Omega_{m}$ from
$\Omega_{m}^{0}$ the contribution of $V(\varphi)$ remains small
for all $z$ in comparison with the cold matter energy density.
More general problem of Lagrangian reconstruction including
non-spatially-flat case has been considered in \cite{Hoyle}.

\subsection{K-essence: F is given, find V}

\indent\indent Now we consider  equations
(\ref{eq12}),(\ref{eq13}) with a kinetic term of the form
\begin{equation}\label{eq17}
F(S)=(a\vert S\vert^{\alpha}+b)^{\beta};
\end{equation}
for definiteness we consider positive $a,b,\alpha,\beta$. The left-hand side of (\ref{eq12}) is a monotonous function, so $S$ and therefore $d\varphi/dz>0$ is uniquely defined from (\ref{eq12}). This gives the monotonous function $\varphi(z)$ up to the additive constant. The potential $V(\varphi)$ also turns out to be a monotonous function. We denote $\gamma=\alpha\beta$.

For $\gamma<1$ the field asymptotic behavior is
\begin{equation*}
\varphi(z)\sim\frac{1}{^{1-\gamma}}(\Omega_{m}^{0}-\Omega_{m})^{1/(2\gamma)}z^{3(1-\gamma)/(2\gamma)}+O(\ref{eq1})
\end{equation*}
as $z\to\infty$,
\begin{equation*}
V(\varphi)\sim\frac{1}{(\Omega_{m}^{0}-\Omega_{m})^{\gamma/(1-\gamma)}}\varphi^{2\gamma/(1-\gamma)},\quad \varphi\to\infty.
\end{equation*}

For $\gamma=1$ the potential has exponential  behavior like
(\ref{eq16}). For $\gamma>1$ the scalar field is bounded
$\varphi(z)\to\varphi_{1}$ for $z\to\infty$, where
$\varphi_{1}<\infty$, and $V(\varphi)\to\infty,
\varphi\to\varphi_{1}$.

\subsection{K-essence: V is given, find F}

\indent\indent First of all we note that in  case of the constant
potential $V(\varphi)=V_{0}$, for any non-trivial dependence
$F(S)$, it follows from (\ref{eq13}) that $S=const$. This is
incompatible with (\ref{eq12}) unless the field is constant, $S=0$
and $\Omega_{m}=\Omega_{m}^{0}$.

Consider the power-law potential
\begin{equation}\label{eq18}
V(\varphi )=A\varphi^{n}.
\end{equation}
Then Eq.(\ref{eq13}) takes on the form
\begin{equation}\label{eq19}
F(S)=\rho_{cr}\left[{(\Omega_{m}^{0}-\Omega_{m})\psi^{n}-(1-\Omega_{m}^{0})}\right],
\end{equation}
where $\psi=\varphi/\alpha,\quad\alpha=[(\Omega_{m}^{0}-\Omega_{m})\rho_{cr}/A]^{1/n}$. After differentiation of (\ref{eq19}) with respect to $z$ and combining the result with (\ref{eq12}) we have
\begin{equation}\label{eq20}
 (1 + z)^{3}\frac{d}{dz}\left\{{\ln\left[{(1+z)h(z)\frac{d\psi }{dz}}\right]}\right\}=n\psi^{n-1}\frac{d\psi }{dz},
\end{equation}

In case of the linear potential ($n = 1$) Eq. (\ref{eq20}) leads to a first order linear differential equation with respect to $(d\psi/dz)^{-1}$ yielding an explicit solution
\begin{equation}\label{eq21}
\psi(z)=C_{1}+\int\limits_{0}^{z}{dz'}\left\{{(1+z')h(z')\left[{C_{2}-\int\limits_{0}^{z'}{\frac{dz''}{(1+z'')^{4}h(z'')}}}\right]}\right\}^{-1},
\end{equation}
$C_{1}$, $C_{2}$ are arbitrary constants. Then
\begin{equation}\label{eq22}
S=\frac{\alpha^{2}H_{0}^{2}}{2}\left[{C_{2}-\int\limits_{0}^{z}{\frac{dz'}{(1+z')^{4}h(z')}}}\right]^{-2}.
\end{equation}
Equations (\ref{eq19}), (\ref{eq22}) define $F$ parametrically.

Denote
\begin{equation*}
 I_{0}=\int\limits_{0}^{\infty}{\frac{dz'}{(1+z')^{4}h(z')}}
\end{equation*}

\begin{description}
 \item[(i)] In case of $C_{2}>I_{0}$ the function $S(z)$ is monotonically increasing on $z\in(-1,\infty)$; it varies within the interval $(0,S_{\max})$, $S_{\max}=(\alpha^{2}H_{0}^{2}/2)\left[{C_{2}-I_{0}}\right]^{-2}$. Therefore $F$ and $\varphi$ may be defined as single-valued functions of $S$ only on $(0,S_{\max})$ and we are free for arbitrary choice of $F(S)$ for $S>S_{\max}$.
\item[(ii)] In case of $C_{2}=I_{0}$ we have $\varphi(z)\sim z^{3}$ and $F(S)\sim S^{1/3}\sim z^{3}$ for $S\to S_{\max}=\infty$.
\item[(iii)] If $0<C_{2}<I_{0}$, then there is a field  singularity $\varphi\sim\ln(z_{1}-z)$, $F(S)\sim\ln(S)$ for some $z\to z_{1}$ and, accordingly, singularity of the energy density. The solution of the system (\ref{eq11}),(\ref{eq19}) cannot be extended for all $z$ in the past. Analogous situation occurs in the future for $C_{2}<0$.
 \end{description}

In a general case $n>0$ it follows from (\ref{eq11})  that in a
regular point of the function $F(S)$ the derivative $d\varphi/dz$
cannot be zero. Equation (\ref{eq20}) may be written as
\begin{equation*}
\frac{(1 + z)^3}{2S}\frac{dS}{dz} = n\psi ^{n - 1}\frac{d\psi }{dz},
\end{equation*}
whence $dS/dz\ne0$; so $S(z)$ and $\varphi(z)$ are  monotonous
functions and there exists a single-valued inverse function $z(S)$
on some interval. Depending of the initial conditions for
(\ref{eq20}) the following cases are possible for solutions of
(\ref{eq20}) for $z>0$:
\begin{description}
 \item[(i)] $S(z)$ is bounded for all $z$;
 \item[(ii)] $S(z)\to\infty$ as $z\to\infty$, in  this case there is a solution having power-law asymptotics $\psi(z)\cong\left({\frac{1}{n}+\frac{1}{2}}\right)^{1/n}z^{3/n}$, $z\to\infty$;
 \item[(iii)] $S(z)\to\infty,\quad \varphi(z)\to\infty$ as $z\to z_{1}$ for some $z_{1}<\infty$. In this case (\ref{eq20}) yields
\begin{equation*}
 (1+z_{1})^{3}\frac{d}{dz}\left\{{\ln\left[{\frac{d\psi}{dz}}\right]}\right\}=\frac{d\psi^{n}}{dz},
\end{equation*}
 \end{description}
where we have omitted the bounded terms that are  not essential
for asymptotical behavior of $\psi$ for $z\to z_{1}$. After
substitution $\psi=(1+z_{1})^{3/n}\xi^{1/n}$ (neglecting an
integration constant) we have
\begin{equation*}
 \ln\left[{\frac{\xi^{(1-n)/n}}{n}\frac{d\xi}{dz}}\right]=\xi,
\end{equation*}
whence
\begin{equation*}
\int\limits_{z}^{\infty}{\xi^{(1-n)/n}e^{-\xi}d\xi}=C_{3}(z_{1}-z),
\end{equation*}
$C_{3}$ is an integration constant.

The asymptotic expansion of the left-hand side integral in powers of $\xi^{-1}$ is
\begin{equation*}
 \int\limits_{z}^\infty{\xi^{(1-n)/n}e^{-\xi }d\xi}=\xi^{(1-n)/n}e^{-\xi}\left({1+\frac{1-n}{n\xi}+\frac{(1-n)(1-2n)}{n^{2}\xi^{2}}+...}\right),
\end{equation*}
whence we get
\begin{equation*}
\psi(z)=\left[{-(1+z_{1})^{3}\ln(z_{1}-z)}\right]^{1/n}\left({1+O\left({\frac{\ln(\ln(z_{1}-z))}{\ln(z_{1}-z)}}\right)}\right).
\end{equation*}

\section{Discussion}

\indent\indent We considered the FRW  cosmological models without
$\Lambda$-term but with the scalar field and the pressureless
matter in case of spatially flat Universe. We study the scalar
field Lagrangians $L=F(S)-V(\varphi)$ yielding the same dependence
$H(z)$ as in the $\Lambda$CDM model, which provides the same
dependence of photometric distance and angular diameter distance
upon redshift. The $H(z)$ dependence restricts the potential
$V(\varphi)$ up to integration constants, provided that the
kinetic term $F(S)$ be given, and vise versa. The exact analytic
expression in parametric form are presented for $V$ if $F(S)\equiv
S$; and for $F$ in case of the linear potential $V(\varphi )$.

If the scalar field is introduced,  then some part of cosmological
density $\Omega_{field}$ will be due to this field. If
$\Omega_{m}^{0}$ (the dark matter content) is derived within the
standard model on account the data on large-scale geometry, then
introduction of the scalar field leads to the relation
$\Omega_{m}+\Omega_{field}=\Omega_{m}^{0}$ and therefore to some
change of the real content of the dark matter. Therefore
$\Omega_{m}$ and/or $\Omega_{field} $ cannot be separated on
account of purely geometric data without additional information
constraining  these parameters separately. In fact the same
dependencies of the Hubble parameter, photometric distance and
angular diameter distance upon redshift may take place for
different relations of $\Omega_{m}$ and $\Omega_{field}$. This
degeneracy can be removed using independent data about
$\Omega_{m}$, which, e.g., are related to the large-scale
structure and/or whole CMB anisotropy spectrum, however in this
case the accuracy of determination of $\Omega_{m}$ will be worse
in comparison with purely $\Lambda$CDM model. It should be pointed
out that this degeneracy is due to the unknown Lagrangian; this
problem does not arise if we specify the appropriate Lagrangian
form up to some parameter set; e.g., in case of a canonical
kinetic term and
$V(\varphi)=\alpha\varphi^{2}+\beta\varphi^{3}+\gamma\varphi^{4}$
the model parameters can be constrained uniquely (within
experimental errors) from the observed $H(z)$ dependence. On the
other hand, it would be highly improbable that realistic
Lagrangian must have exactly one of the forms describes in Section
3 so as to mimic the $\Lambda$CDM model. However, a rigorous
approach must rule out such possibility as well.

The examples considered above when  the kinetic term is specified
(subsections 3.1, 3.2) show the infinite behavior of the field in
the past though the relative content of the field energy density
remains typically of the same order as at the present epoch
($\sim\Omega_{m}^{0}-\Omega_{m})$. The other set of examples
(subsection 3.3) shows singular behavior for finite $z$ and thus
suggests possibility of new physical situation (phase
transitions?) in the early Universe.

At the end we note that though the  above considerations deal with
the case of spatially flat Universe and pressureless matter, main
qualitative aspects (degeneracy in determination of $\Omega_{m}$
and existence of singular behavior in the past) will remain in a
more general case dealing with more general dark matter equation
of state or $H(z)$ dependence.

\section*{Acknowledgements}
\indent \indent This work has been  supported in part by the
program "Cosmomicrophysics" of National Academy of Sciences of
Ukraine.

\end{document}